\documentclass[twoside]{LCWS11}
\usepackage[latin1]{inputenc}
\usepackage{graphicx,color}
\usepackage{wrapfig,rotating}
\usepackage{amssymb,amsmath,array}
\usepackage{cite}
\begin{document}
\title{
Indirect Sensitivity to Heavy Z$'$ Bosons\\ at a Multi-TeV $e^+e^-$ Collider} 
\author{Marco Battaglia$^{1,2,3}$, 
Francesco Coradeschi$^{4,5}$, 
Stefania De Curtis$^5$
and Daniele Dominici$^{4,5}$ 
\vspace{.3cm}\\
1- University of California at Santa Cruz - Santa Cruz Institute of Particle Physics,  \\
Santa Cruz, CA 95064 - USA\\
2- Lawrence Berkeley National Laboratory, Berkeley, CA 94720 - USA\\
3- CERN, CH-1211 Geneva - Switzerland\\
4- Universit\'a degli Studi - Dipartimento di Fisica e Astronomia, I-50019 Firenze, Italy\\
5- Istituto Nazionale di Fisica Nucleare, Sezione di Firenze, I-50019 Firenze, Italy\\
}

\maketitle

\begin{abstract}
We compare the phenomenology of two models, the so-called ``minimal $Z'$'' and an effective 
model for a SM-like Higgs realised as a composite state of a new strong interaction, at a 
multi-TeV linear collider in the hypothesis that the new physics is at a scale beyond the 
direct reach of the machine.
\end{abstract}

\section{Introduction}

A variety of Standard Model extensions, such as grand unified models, 
strings and branes, models of extra dimensions and models utilising alternative 
schemes of electro-weak symmetry breaking, usually as an unbroken remnant of a larger 
gauge symmetry, include extra gauge bosons. Current limits from direct searches 
at the LHC constrain their mass above $\sim$1.1 - 1.7~TeV, depending on their 
couplings~\cite{ATLAS,CMS}. The LHC is expected to explore the region of mass up to several TeV. 

In this study, we discuss the sensitivity of a high-energy $e^+e^-$ collider to different models, 
containing extra neutral $Z'$ bosons, away from its centre-of-mass energy, $\sqrt{s}$ . First, we 
briefly comment on the sensitivity to states at masses below $\sqrt{s}$ with an ``auto-scan'' 
technique, which makes use of the luminosity spectrum tail due to radiation. Then, we analyse 
electro-weak precision observables in $e^- e^+ \to f \bar{f}$ processes for two reference models with 
$M_{Z'} > \sqrt{s}$. The first model is the so-called ``minimal $Z'$'' \cite{Appelquist:2002mw,Salvioni:2009mt}, 
where new physics is at a very high energy scale and manifests itself at the TeV scale through a single 
$Z'$ boson. In the second model~\cite{Contino:2006nn}, 
the Higgs field and other fields, including three $Z'$s, are realised as composite states from a 
strong interaction at the TeV scale. In the following we refer to this as the Effective Composite Higgs 
Model (ECHM). These two models depict different physical situations and each represents one of the 
simplest realisations of the corresponding scenario, so they are well-suited to be used to test the 
collider sensitivity.

\section{Simulation and Data Analysis}

In the SM, $e^- e^+ \to f \bar{f}$ processes can be fully parametrised in terms of four helicity 
amplitudes, which can be in turn determined by measuring four observables: the total production 
cross section, $\sigma_{f \bar f}$, the forward-backward asymmetry, $A_{FB}$, the left-right 
asymmetry, $A_{LR}$, and the polarised forward-backward asymmetry, $A^{pol}_{FB}$. These observables 
still characterise the $e^- e^+ \to f \bar{f}$ process if $Z'$ bosons are the only new neutral states. 
In fact, in the case of a single $Z'$ of known mass they can be used to determine the new vector 
couplings~\cite{Riemann:2001bb}. 

The anticipated experimental accuracy on these electro-weak observables for the $e^+e^- \to f \bar f$ 
($f$ = $\mu$, $b$ and $t$) processes are determined  
from the analysis of fully simulated and reconstructed $e^+e^- \to f \bar f$ events at $\sqrt{s}$=3~TeV 
using the ILD detector model~\cite{ild}, modified for physics at CLIC. Beamstrahlung effects are 
taken into account but no machine-induced-backgrounds are overlayed. 
For polarised observables we assume 80\% and 60\% polarisation for the $e^-$ and $e^+$ 
beam, respectively. $b$-tagging is based on the response of the vertexing variables of the {\tt ZVTOP} 
algorithm~\cite{Bailey:2009ui}. These characterise the kinematics and topology of the secondary system 
in a jet. They are supplemented by the corresponding kinematic observables for the secondary system 
based on particle impact parameters instead of topological vertexing, when the {\tt ZVTOP} 
algorithm does not return any secondary vertex. This procedure allows us to increase the efficiency for 
$b$ jets at the higher end of the kinematic spectrum, which is particularly important in this analysis. 
Tagging observables are combined into a single discriminating variable using the boosted decision tree 
procedure in the {\tt TMVA} package~\cite{Hocker:2007ht}. For this analysis we choose a working point 
corresponding to a full energy $b \bar b$ event tagging efficiency of 0.68. For $t \bar t$ tagging, events 
passing the $b$-tagging criteria are reconstructed as two-jet events and at least one of the jets is 
required to be compatible with the top quark mass. This gives an efficiency for full energy $t \bar t$ 
events of 0.55. 
Quark charge is determined using the lepton charge in $b$ and $t$ semileptonic decays, which is robust 
against the effect of machine-induced backgrounds, contrary to the case of jet or vertex charge techniques. 
In particular, for $t \bar t$ events we tag the top production using the hadronic decay of one top quark 
and determine the charge using the lepton from the $W^{\pm} \to \ell^{\pm} \nu$ decay in the opposite 
hemisphere.

\begin{figure}[h]
\begin{center}
\includegraphics[width=0.45\textwidth]{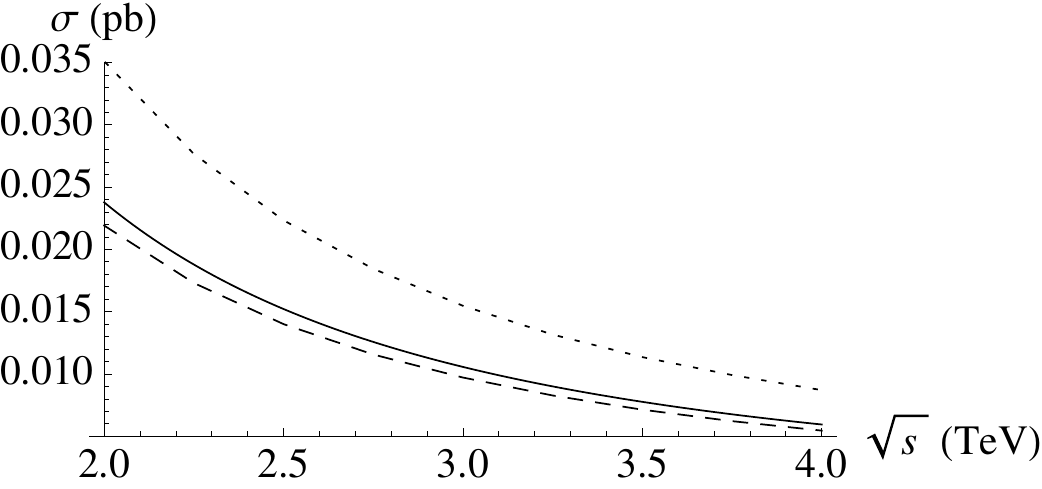}
\caption{Cross section for the $e^+ e^- \to \mu^+ \mu^-$ in the SM at the Born level 
(solid line), with ISR and beamstrahlung and no cuts (dotted line), and with ISR, beamstrahlung and 
the cut $E_{f,\bar{f}}>0.8 \, E_{beam}$ (dashed line). The cut selects final state fermions which 
did not experience significant energy radiation, bringing the cross-section back within $5-10 \%$ of its 
Born-level value.}
\label{fig:isr}
\end{center}
\end{figure}

The electro-weak observables for the $Z'$ models and the SM are computed using CalcHEP \cite{calchep}. 
The model files for the ECHM model are generated from the Lagrangian using the 
{\tt FeynRules} \cite{Christensen:2008py} package in Mathematica~\cite{mathematics}, and its couplings 
calculated by implementing an external C library to obtain a numerical diagonalisation of the mass matrices. 
CalcHEP matrix elements are obtained at tree level and corrections from Initial State Radiation (ISR) 
and beamstrahlung are added. ISR is implemented using the formalism of~\cite{Jadach:1988gb}. 
These calculations include a few event selection cuts. First, a cut on the polar angle of the final state fermions, 
$|\cos \theta|< 0.9$, ensures their observability in the detector. Then, a cut on the final state energy, 
$E_{f,\bar{f}}>0.8 \, E_{beam}$, selects high energy events. This removes the effect of radiation and brings 
the  visible cross-section for the process down to its Born cross section value as shown in Figure~\ref{fig:isr}.

\section{$Z'$ models at $\sqrt{s}$ = 3~TeV}

\subsection{Direct sensitivity}

If a new neutral resonance were to be observed at the LHC it would become extraordinarily 
interesting to produce it in lepton collisions and accurately determine its properties and nature.
A multi-TeV $e^+e^-$ collider, such as CLIC, is very well suited for such a study.
By precisely tuning the beam energies to perform a detailed resonance scan the parameters
of the resonance can be extracted with high accuracy~\cite{Battaglia:2000am,Accomando:2004sz}. 
Operating CLIC at its full design energy of 3~TeV it will also be possible to search for new 
resonances coupled to $e^+e^-$ and perform a first determination of their mass and width using 
the beamstrahlung and ISR tail through an ``auto-scan'' without changing its beam energy. 
\begin{figure}[hb!]
\begin{center}
\includegraphics[width=0.42\textwidth]{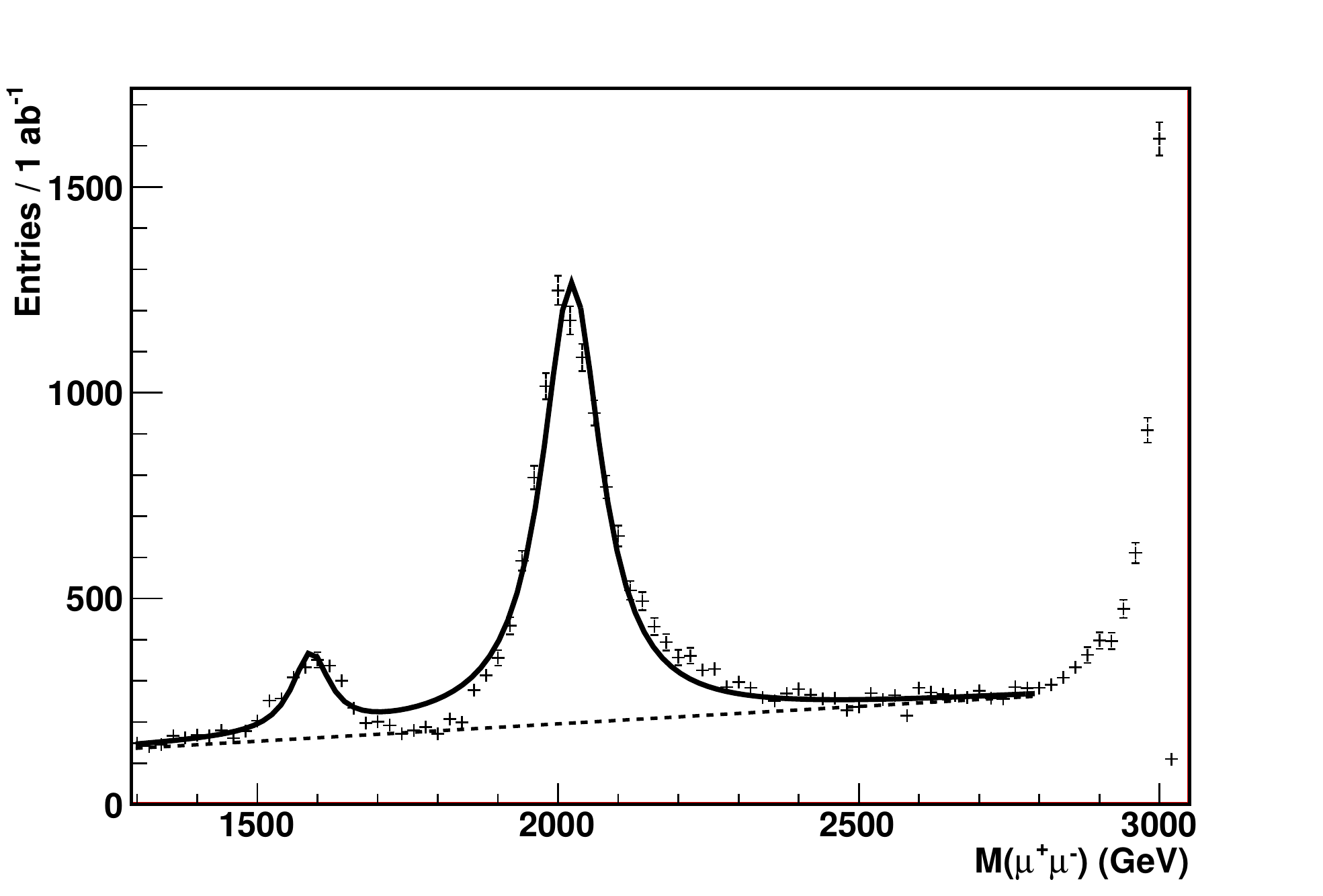} \\
\caption{Observation of new gauge boson resonances in the $\mu^+\mu^-$ channel by auto-scan with 
1~ab$^{-1}$ of data at 3~TeV . The two resonances are the $Z_{1,2}$ predicted by the 4-site model, 
an effective scheme for a 5-D Higgs-less model~\cite{Accomando:2008jh}.}
\label{fig:scan}
\end{center}
\end{figure}
An example is given in Figure~\ref{fig:scan} where the  invariant mass of $\mu^+\mu^-$ pairs in 
the $e^+e^- \to \mu^+\mu^-$ process is shown for the case of two new neutral gauge bosons arising 
from  models with extra dimensions. The masses can be determined with 1~ab$^{-1}$ data to 
(1588.7$\pm$3.5)~GeV and (2022.6$\pm$1.2)~GeV, i.e.\ with a statistical accuracy of just a few 
per-mil, or better, by operating the collider always at the maximum $\sqrt{s}$ energy.

\subsection{Indirect sensitivity}

\begin{figure}[h!]
\includegraphics[width=0.42\textwidth]{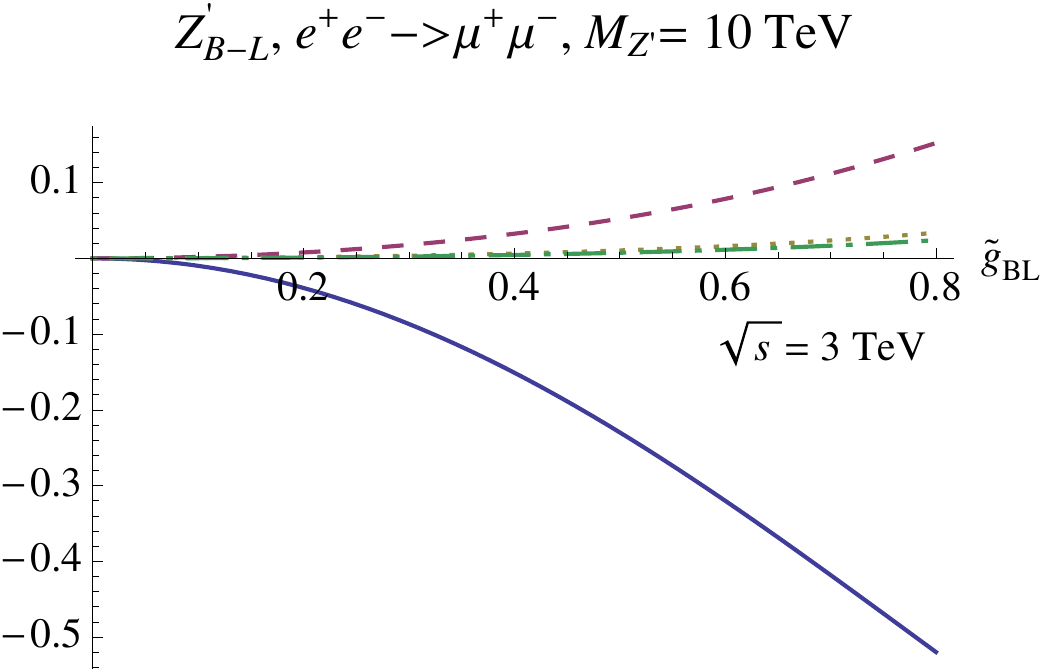}~~~~
\includegraphics[width=0.48\textwidth]{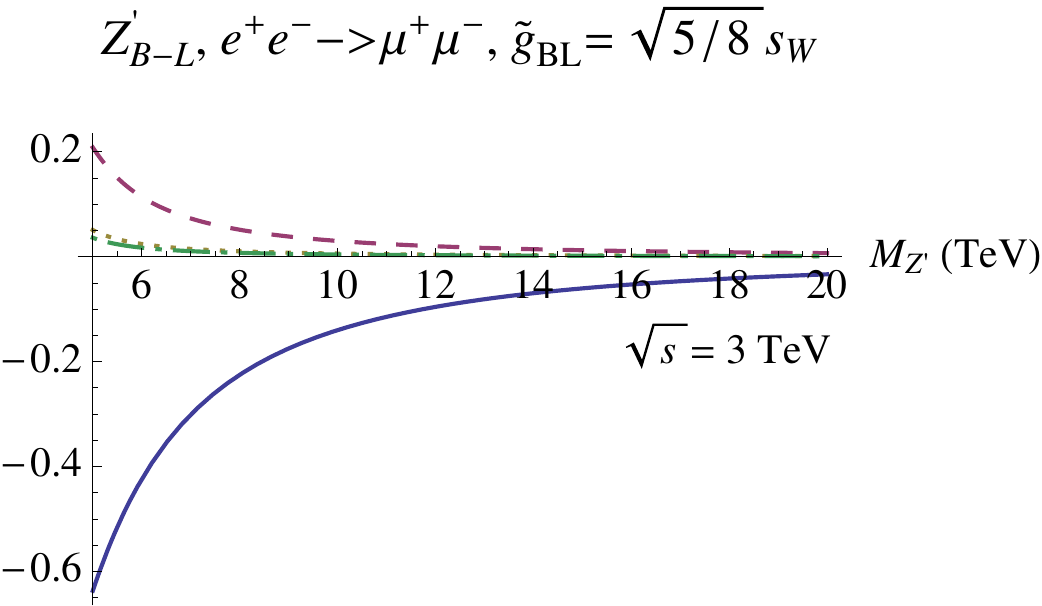}
\caption{Sensitivity of $e^+e^- \to \mu^+ \mu^-$ in the $Z'$ minimal  model, for parameters corresponding 
to the B-L model, at
$\sqrt{s}$=3~TeV. The deviations the $\sigma_{\mu \mu}$, $A_{FB}$, $A_{LR}$ and  $A_{FB}^{pol}$ 
observables to the SM predictions are shown in the left panel as a function of $\tilde g_{BL}$ 
for $M_{Z'}$=10~TeV and in the right panel as function of $M_{Z'}$ for a fixed value of 
$\tilde g_{BL}$. The continuous line represents the deviation $(\sigma - \sigma^{SM})/\sigma^{SM}$, 
the dashed line $A_{FB} - A_{FB}^{SM}$, the dotted line $A_{LR} - A_{LR}^{SM}$ 
and the dot-dashed line $A_{FB}^{pol} - A_{FB}^{pol~SM}$.} 
\label{fig:Zmin}
\end{figure}

In case no signal is observed at the LHC, a multi-TeV $e^+e^-$ could still obtain 
essential information on extra gauge bosons by a precision study of electro-weak observables, 
sensitive to the effects of new particles at mass scales well above $\sqrt{s}$. Here, we consider 
two different scenarios with heavy $Z'$ bosons. Besides the so-called Sequential Standard Model (SSM) 
and the $E_6$-inspired models, which have been already extensively studied~\cite{Battaglia:2002sr,Accomando:2004sz}, 
there is a more general and model-independent parametrisation of a $Z'$ boson and its couplings, as 
proposed in~\cite{Appelquist:2002mw}, generally referred to as "minimal" $Z'$ model. Its phenomenology 
at LHC has been recently studied in detail~\cite{Salvioni:2009mt}. 
The basic assumption in the model description is the presence of a single new vector boson state 
with a mass of order TeV plus the \emph{minimal amount} of extra non-SM fields needed to make the 
model renormalisable and free of anomalies. 
Before mixing, the coupling of the $Z'$ to fermions can be written as:
\begin{equation}
\label{ZpminLint}
\mathcal{L}_{int}^{Z'} = i g_{Z} Z'_\mu \bar{f} \gamma^\mu (\tilde{g}_Y Y + \tilde{g}_{BL} (B-L)) f,
\end{equation}
where $g_Z$ is the standard $Z$ coupling, and $Y$, $B$ and $L$ are the usual hypercharge,
baryon and  lepton numbers.
The effects from the most general kinetic and mass mixing can be  described in terms
of two independent couplings of the $Z'$ to fermions, $\tilde{g}_Y$ and $\tilde{g}_{BL}$, 
which induce $Z-Z'$ mixing. Several of the well-known $Z'$ models considered earlier on 
can be incorporated in this framework by fixing the ratio $\tilde{g}_Y / \tilde{g}_{BL}$.
The deviations of the electro-weak observables in the  $e^+e^- \to \mu^+ \mu^-$ channel are shown 
in Figure~\ref{fig:Zmin}, for the case of the B-L model~\cite{Langacker:2008yv}. 
These are comparable in size 
for the different final state fermions considered here.

\begin{figure}[h!]
\includegraphics[width=0.48\textwidth]{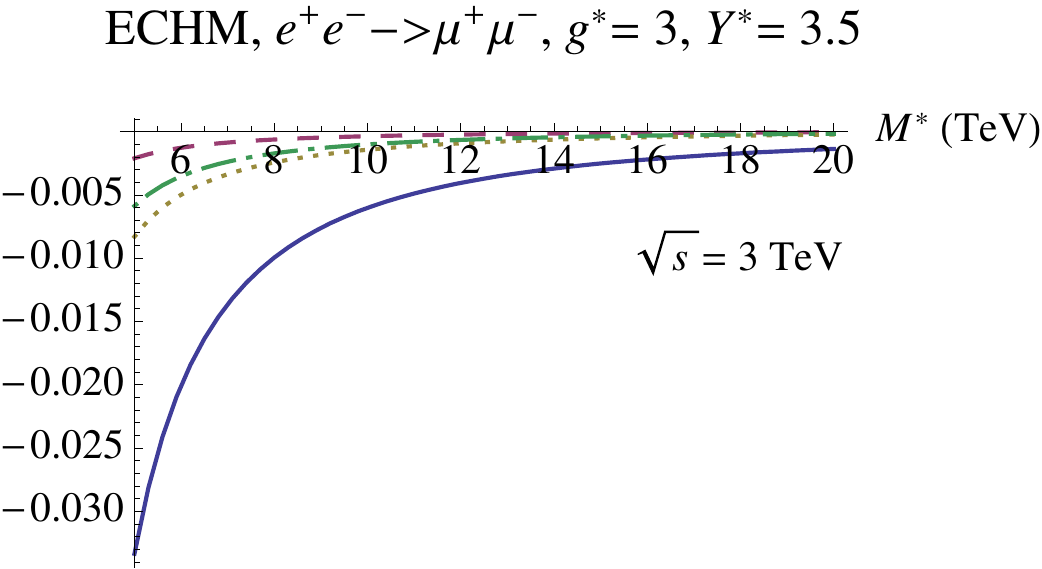}~~~~
\includegraphics[width=0.46\textwidth]{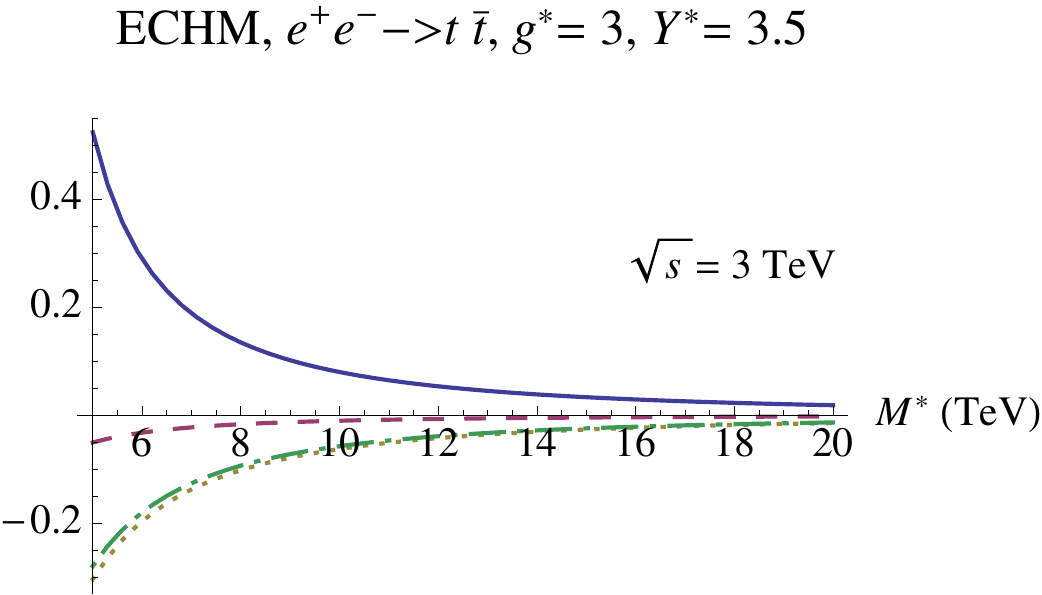}
\caption{Sensitivity of the $\sigma_{\mu \mu}$, $A_{FB}$, $A_{LR}$ and  $A_{FB}^{pol}$ 
observables in the ECHM  model at $\sqrt{s}$=3~TeV. The deviations w.r.t.\ the SM predictions are shown 
as a function of  $M_{Z'}$ for a fixed values of the couplings for $e^+e^- \to \mu^+ \mu^-$ in the left 
and for $e^+e^- \to t \bar t$ in the right panel. The convention on the symbols is the same as in 
Figure~\ref{fig:Zmin}.} 
\label{fig:ECHM}
\end{figure}

The ECHM model introduced in~\cite{Contino:2006nn} represents a qualitatively different scenario, 
where third generation fermions play a special role. The model can be described as a "maximally 
deconstructed", i.e.\ with the extra dimension discretised down to just two sites, version of 
the so-called "RS custodial" 5-dimensional model, first studied in~\cite{Agashe:2003zs}, 
related via the ADS/CFT correspondence to the scenario of partial compositeness of 
the SM. This model aims at explaining the fermion mass hierarchy and to stabilise the Higgs sector.
It describes the SM fields and their first KK composite excitations as a result of two sectors, 
elementary and composite, which are mixed. In the neutral sector, there are three heavy $Z'$s. 
Their couplings are controlled by composite-elementary mixing angles, which are generation-dependent. 
In this phenomenological analysis we have assume universal new vector boson mass parameter $M^*$ and 
composite gauge coupling $g^*$. In order to study the modifications of the standard four-fermion operators 
from $Z'$ exchanges, we also assume that the composite fermions have an universal mass scale, 
$m^*$, which is taken to be greater than $M^*$ by fixing $m^* = 1.5 M^*$. This ensures that $M^*$ is the 
only relevant mass scale in study of the electro-weak observables.
In the Yukawa/fermion mixing sector, we have assumed full $t_R$ compositeness~\cite{Contino:2006nn} while
fermions other than the quarks of the $3^{rd}$ generation are taken to be mostly elementary. 
Our phenomenological analysis has three free parameters: $M^*$, $g^*$, and $Y_{*U33}$, where the latter 
is the Yukawa coupling in the composite fermion sector related to the top mass.
Due to the relatively strong experimental constraints on the  $Z b_L b_L$ coupling, the $b$ quark couplings 
to the three heavy neutral vectors must be taken to be lower than those of the SM, while those of the 
$t$ are enhanced. The main signature of this model is in the large deviations of the top sector 
observables from their SM expectations. This is shown in Figure~\ref{fig:ECHM}, where the deviations of the 
electro-weak observables in the $\mu^+ \mu^-$ and $t \bar t$ final states  from their SM values are shown as 
a function of $M^*$.

\section{Results}

\begin{figure}[h!]
\begin{center}
\begin{tabular}{cc}
\includegraphics[width=0.42\textwidth]{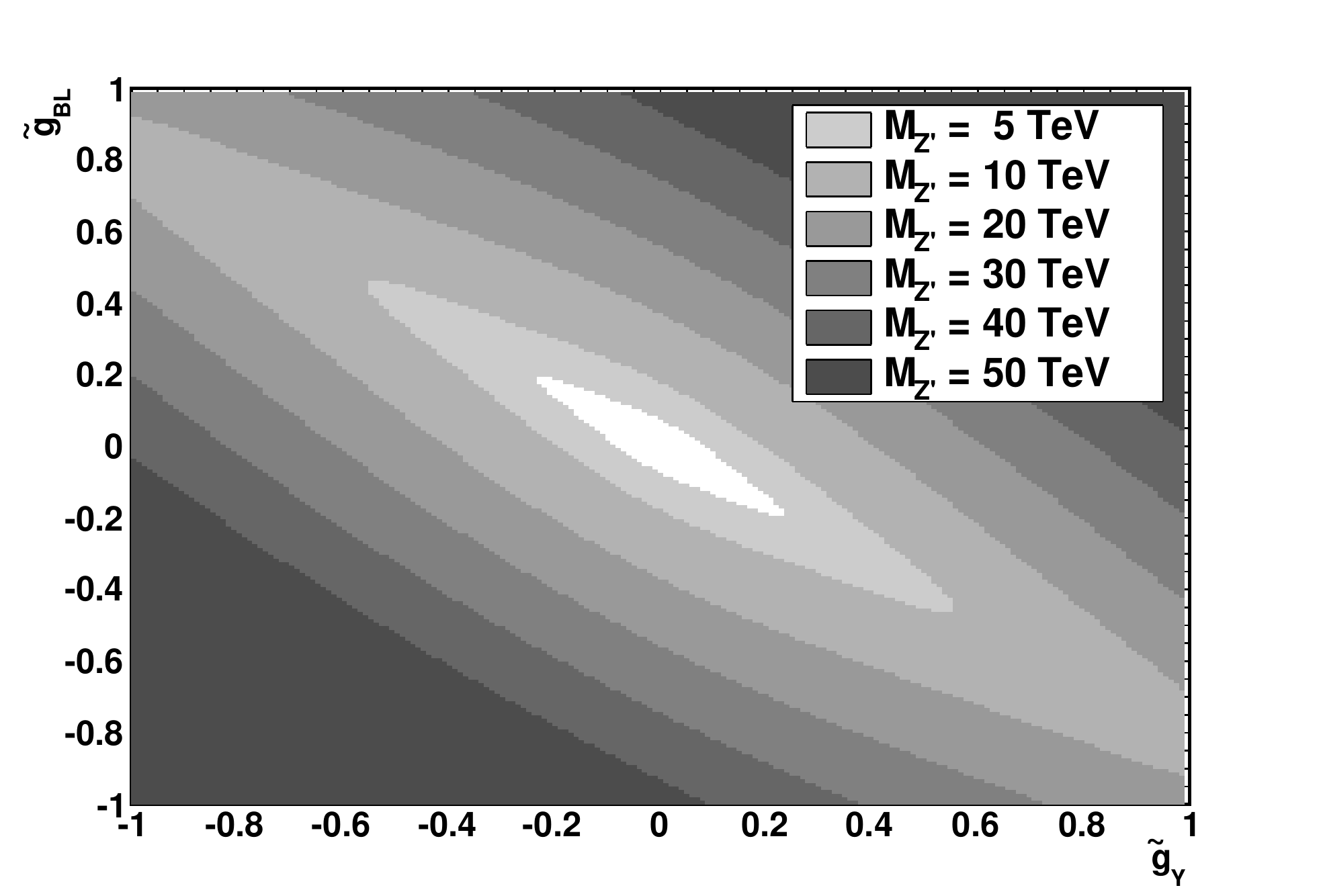} &
\includegraphics[width=0.42\textwidth]{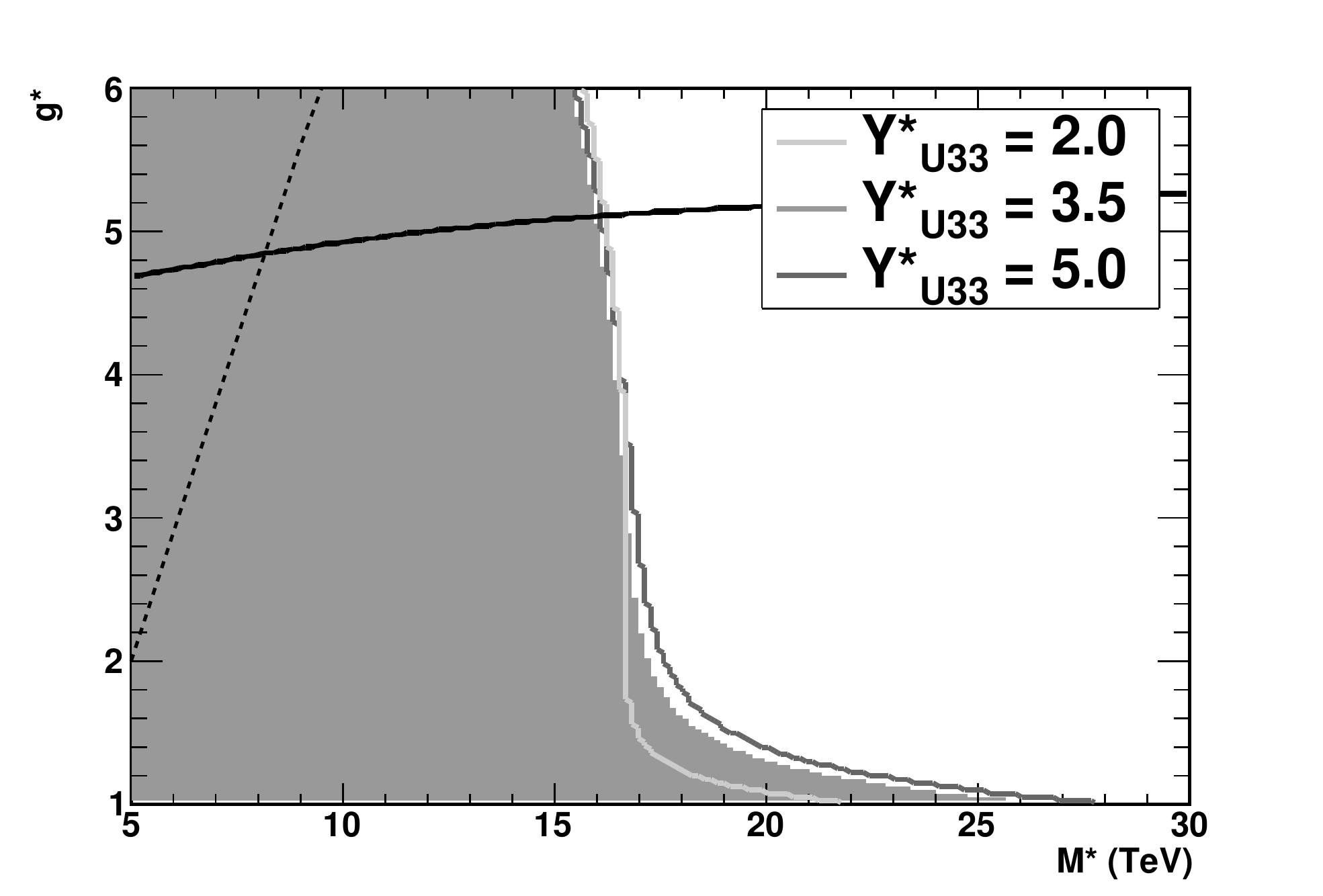} \\
\end{tabular}
\caption{Left: Sensitivity to the $Z^{'}$ minimal model in the $\tilde{g}_{BL}$ 
vs. $\tilde{g}_Y$ plane for various values of $M_{Z'}$. Right: Sensitivity to the ECHM in the 
($M*$, $g*$) plane for different values of $Y*_{U33}$. The region above the continuous line has 
the broader resonance with $\Gamma >$ 0.5 $M$ and our perturbative calculations cannot be trusted. 
The region above the dashed line is excluded by present electro-weak data for $Y*_{U33}$ = 2 but allowed 
for larger values. In both plots we assume $\sqrt{s}$ = 3~TeV with 2~ab$^{-1}$ of integrated luminosity 
and polarised beams (80\% for $e^-$ and 60\% for $e^+$).}
\label{fig:ZpminEx}
\end{center}
\end{figure}

We estimate the sensitivity of our observable to these $Z^{'}$ models in terms of the region in parameter phase space 
where the electro-weak observables have values incompatible with their SM predictions. 
Here we use nine of the electro-weak observables, i.e.\ production cross section, $\sigma_{f \bar f}$, forward-backward 
asymmetry, $A_{FB}$, and left-right asymmetry, $A_{LR}$, for the $\mu^+ \mu^-$, $b \bar b$ and $t \bar t$ final states. 
We perform flat scans of the parameter space of each model on a grid and for each scan point we compute the pulls, 
given by the differences between the actual model point values from their SM values normalised to the measurement 
accuracy estimated from simulation. The sensitivity to a model is defined as the region of 
parameters where the $\chi^2$ probability for the nine electro-weak observables to be compatible with the 
SM prediction is below 0.05. As highlighted by the previous studies, $e^+e^-$ collider data are generally sensitive to 
mass scales which larger than the $\sqrt{s}$ value by a factor varying from a few times to an order of magnitude, which 
pushes the typical sensitivity of a multi-TeV collider well beyond the direct accessibility of any realistic particle 
collider. Results for the $Z^{'}$ minimal model and the ECHM model are shown in Figure~\ref{fig:ZpminEx}. They confirm 
that level of sensitivity for the two models studied here with a range of sensitivity varying between few TeV to 30~TeV 
and beyond, depending on the couplings for the  $Z^{'}$ minimal model and around 16~TeV for the ECHM model. In this 
case there is a weak dependence on the values of the other parameters, except for $g^* \simeq$ 1, where the mass 
sensitivity exceeds 20~TeV.

\begin{footnotesize}

\end{footnotesize}

\end{document}